\documentclass[sigconf]{acmart}

\usepackage[acronym,nohypertypes={acronym},shortcuts]{glossaries}
\usepackage{amsfonts}
\usepackage{listings}
\usepackage{color}
\usepackage{stfloats}
\usepackage{soul}
\usepackage{diagbox}

\AtBeginDocument{%
  \providecommand\BibTeX{{%
    \normalfont B\kern-0.5em{\scshape i\kern-0.25em b}\kern-0.8em\TeX}}}

\setcopyright{acmcopyright}
\copyrightyear{2024}
\acmYear{2024}
\acmDOI{10.1145/xxxxxxx.xxxxxxx}

\acmConference[WiSec '24]{WiSec '24: 17th ACM Conference on Security and Privacy in Wireless and Mobile Networks}{May 27--May 30, 2024}{Seoul, Korea}
\acmBooktitle{WiSec '24: 17th ACM Conference on Security and Privacy in Wireless and Mobile Networks, May 27--May 30, 2024, Seoul, Korea}
\acmPrice{15.00}
\acmISBN{978-1-4503-XXXX-X/21/06}

\begin{document}

\title{DynamiQS: Quantum Secure Authentication for Dynamic Charging of Electric Vehicles}


\author{Tommaso~Bianchi}
\orcid{0000-0001-8192-5117}
\affiliation{%
  \institution{University of Padua}
  \city{Padua}
  \country{Italy}}
\email{tommaso.bianchi@phd.unipd.it}

\author{Alessandro~Brighente}
\orcid{0000-0001-6138-2995}
\affiliation{%
  \institution{University of Padua}
  \city{Padua}
  \country{Italy}}
\email{alessandro.brighente@unipd.it}

\author{Mauro~Conti}
\orcid{0000-0002-3612-1934}
\affiliation{%
  \institution{University of Padua}
  \city{Padua}
  \country{Italy}
}
\email{mauro.conti@unipd.it}

\renewcommand{\shortauthors}{Bianchi Tommaso, et al.}

\begin{abstract}
Dynamic Wireless Power Transfer (DWPT) is a novel technology that allows charging an electric vehicle while driving thanks to a dedicated road infrastructure. DWPT's capabilities in automatically establishing charging sessions and billing without users' intervention make it prone to cybersecurity attacks. Hence, security is essential in preventing fraud, impersonation, and user tracking. To this aim, researchers proposed different solutions for authenticating users. However, recent advancements in quantum computing jeopardize classical public key cryptography, making currently existing solutions in DWPT authentication nonviable. 
To avoid the resource burden imposed by technology upgrades, it is essential to develop post-quantum-resistant solutions.

In this paper, we propose DynamiQS, the first post-quantum secure authentication protocol for dynamic wireless charging. DynamiQS is privacy-preserving and secure against attacks on the DWPT. 
We leverage an Identity-Based Encryption with Lattices in the Ring Learning With Error framework. Furthermore, we show the possibility of using DynamiQS in a real environment, leveraging the results of cryptographic computation on real constrained devices and simulations. DynamiQS reaches a total time cost of around 281 ms, which is practicable in dynamic charging settings (car and charging infrastructure).
\end{abstract}



\maketitle

\newacronym{ev}{EV}{Electric Vehicle}
\newacronym{kaist}{KAIST}{Korea Advanced Institute of Science \& Technology}
\newacronym{olev}{OLEV}{On-Road Dynamic Wireless Charging Technology}
\newacronym{dwpt}{DWPT}{Dynamic Wirelss Power Transfer}
\newacronym{cp}{CP}{Charging Plate}
\newacronym{nist}{NIST}{National Institute of of Standards and Technology}
\newacronym{rlwe}{RLWE}{Ring Learning With Errors}
\newacronym{obu}{OBU}{On-Board Unit}
\newacronym{pq}{PQ}{Post-Quantum}
\newacronym{ibe}{IBE}{Identity-Based Encryption}
\newacronym{csp}{CSP}{Charging Service Provider}
\newacronym{pkg}{PKG}{Private Key Generator}\
\newacronym{ca}{CA}{Certificate Authority}
\newacronym{ra}{RA}{Registration Authority}
\newacronym{rsu}{RSU}{Road-Side Unit}
\newacronym{ccs}{CCS}{Company Charger Server}
\newacronym{fs}{FS}{Fog Server}
\newacronym{puf}{PUF}{Physical Unclonable Function}
\newacronym{cspa}{CSPA}{Charging Service Provider Authority}
\newacronym{iot}{IoT}{Internet-of-Things}
\newacronym{dsrc}{DSRC}{Dedicated Short-Range Communication}
\newacronym{csma}{CSMA}{Carrier-Sense Multiple Access}
\newacronym{pq}{PQ}{Post-Quantum}

\section{Introduction}\label{sec:intro}
The run to the \acrfull{ev} transition demands the development of charging technologies and infrastructures that can satisfy the users' needs in terms of fast charging services. Currently, \acrshort{ev} drivers can charge their cars only in dedicated stations, with the occurrence of often stopping their trip and leading to the \textit{range anxiety} phenomenon~\cite{range-anxiety}, i.e., a driver fearing running out of battery with an \acrshort{ev}. Stopping for charging may require a considerably long time in case of depleted battery~\cite{pareek}, thus causing delay or downtime during the trip. 
To solve this problem, researchers and engineers are developing infrastructures that allow charging the \acrshort{ev} while driving, starting from the work of Suh et al.~\cite{suh-kim} at \acrfull{kaist}, in which they present \acrshort{olev}: \acrlong{olev}. 
Dynamic wireless charging represents a new charging technology leveraging magnetic induction to transmit power to \acrshort{ev} via wireless power transfer enabled by a set of \acp{cp} under the road~\cite{panchal}. 

This new system eliminates the need to use cables and plug the car in, thus providing opportunities to avoid stopping for a charge. 
Dynamic wireless charging, also called \acrfull{dwpt}, drifted the research towards identifying infrastructures, hardware components, and protocols to make it safe and secure for future customers~\cite{hutchinson}.
Indeed, the evolution of \ac{dwpt} brings new security challenges that can threaten the users if not carefully considered during the development process.  Indeed, attackers can exploit free-riding~\cite{free-riding} and Double-spending attacks~\cite{double-spending}, in which a malicious user utilizes another customer's credentials to get a free charge or re-uses an expired permit. Furthermore, attackers can track customers along different charging processes and undermine the location privacy and identity~\cite{qevsec}. 

To face these novel attacks, researchers proposed various cryptographic authentication protocols. These solutions need to account for the peculiarities of the infrastructure, which involves the pads under the road, the charging process, time and computation constraints, and the security implications derived by the system itself (e.g., identity and location privacy).
The first proposals used simple operations such as XOR and hashing to hide the sensible content to overcome the constraints on the devices, usually equipped with low power and low computation capabilities \cite{dwpt-survey}. 
Recent protocols adopt a public key infrastructure scheme to conceal information under secure operations and manage the customers' credentials~\cite{dwpt-survey}. However, in recent years, the advances in quantum computing are threatening the security of classical public key cryptography solutions.
Indeed, Shor et al.~\cite{Shor_1997} showed that it is possible to break the security of public-key cryptography, and other secure schemes need larger key size~\cite{nist-pq-report}, with the drawback of increased computational, storage, and communication needs. \textbf{As of today, none of the currently existing authentication protocols for \ac{dwpt} can resist attacks from quantum computers}. A problem encountered by Yoshizawa et al.~\cite{pq-v2x} is that the migration to the \acrshort{pq} algorithms requires a transition in which legacy and new schemes coexist. These critical scenarios are especially true in the Vehicle to Everything (V2X) systems, where devices and cars have a long life span and slow adaptation~\cite{pq-v2x}. In fact, the adoption of new technology can take more than 10 years to reach the totality of the circulating vehicle shares~\cite{hula2014analysis, bharadwaj2015technological}, or even worse, not reach the 100\% of cars in production~\cite{hula2014analysis}. Considering that the migration to \acrfull{pq} technologies can be complex for well-established infrastructures, it is fundamental to develop \ac{pq} resistant solutions well in advance.

In this paper, we propose DynamiQS (Dynamic charging Quantum Secure authentication), the first \acrshort{pq} secure identity-based authentication scheme for \ac{dwpt}. Our scheme provides a solution to the advent of quantum computing and the security issues for \acp{ev} charging. We allow direct deployment of a secure framework that considers the users' privacy needs regarding location and identity. The protocol guarantees the customer's anonymity while providing security. We build it on top of proven-to-be-secure mathematical tools that resist quantum computing, such as NTRU Lattices with \acrfull{rlwe} problem. Additionally, we present a network model suitable for the technology and the use of \acrshort{pq} cryptography. The performance evaluation confirms the possibility of deploying it on constrained devices, such as the hardware in the \acrshort{dwpt} infrastructure. Simulations support the theoretical results of a time in the order of around 300 milliseconds. This time is also affordable at high vehicle speed, thanks to the model provided.
To summarize, our contributions are the following:
\begin{itemize}
    \item We propose DynamiQS, the first \acrlong{pq} resistant authentication protocol for \acrshort{dwpt} based on~\cite{ducas} \acrfull{ibe} over NTRU Lattices using \acrshort{rlwe}. The scheme is secure against different attacks. In particular, it preserves the privacy of the customer, avoiding location tracking and identity discovery;
    \item We provide a security formal analysis, along with the use of Scyther tool~\cite{scyther} for validation;
    \item Our protocol is doable for constrained devices, and we prove its applicability on a real scenario leveraging simulations and the work of Güneysu et al.~\cite{guneysu}.
\end{itemize}
Additionally, we adopt a different system scheme with respect to the State-of-The-Art protocols. It is more suitable for real implementation and meets the billing infrastructure needs.

Finally, the paper is structured as follows: Section \ref{sec:background} provides the background about \acrshort{dwpt} technology and authentication schemes. In Section \ref{sec:models} we described our network and adversary models that we adopted to develop this protocol. In Section \ref{sec:protocol} the protocol is explained in its different parts, while in Section \ref{sec:security} and Section \ref{sec:performance}, we provide the security and performance analysis. In the end, we discuss our results in Section \ref{sec:discussion}, we present the related work in Section \ref{sec:related}, and the conclusions in Section \ref{sec:conclusion}.

\section{Background}\label{sec:background}
We describe the background of \acrshort{dwpt} technology and the security requirements that it needs in Section \ref{subsec:dwpt}. 
Sections \ref{subsec:ibe} and \ref{subsec:pq_math} explain the Identity-Based Public Key Infrastructure, and \acrshort{pq} mathematical artifacts for this work, respectively.

\subsection{Dynamic Charging for Electric Vehicles}\label{subsec:dwpt}
\begin{figure}[!htb]
  \centering
  \includegraphics[width=0.85\linewidth]{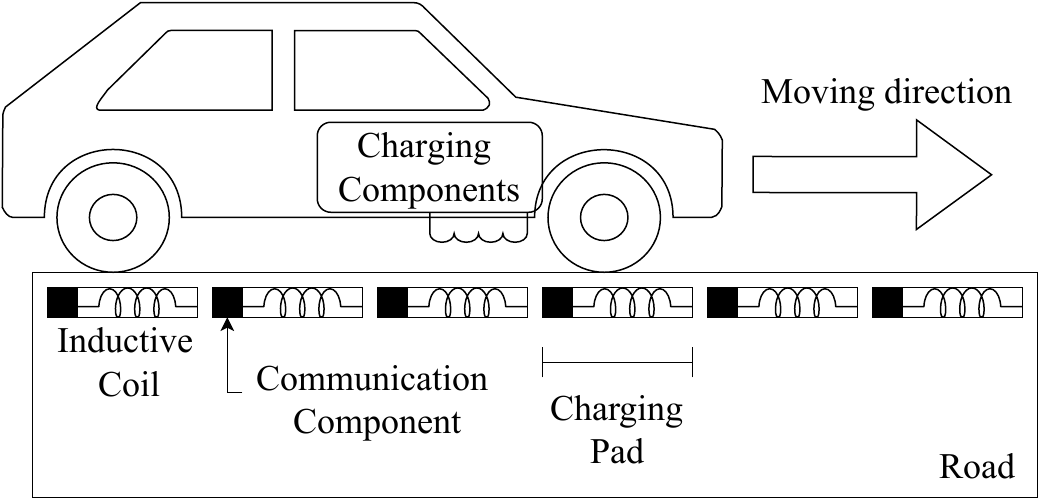}
  \caption{Schematic representation of \acrshort{dwpt} for \acrshort{ev} charging process. The dedicated components on the vehicle receive power while moving over the coils under the road through magnetic induction.}
  \label{fig:dwpt_scheme}
\end{figure}

The \acrshort{dwpt} system is the new frontier for charging an \ac{ev} while driving.
\acrshort{cp}s placed under the road make dynamic charging possible. \acrshort{cp}s form a lane that, when passed over by a car, exchange power with the coils placed on the \acrshort{cp} and on board the vehicle itself~\cite{miller}. 
Figure \ref{fig:dwpt_scheme} represents a \acrshort{dwpt} high-level overview. 
Generally, the vehicle is equipped with a charging component, composed of the \acrfull{obu}, a piece of hardware dedicated to the secure communication and storage of variables and cryptographic keys. 
The \acrshort{cp} has two parts: the inductive coil for power transmission, and a communication element. 
This technology brings new research problems related to the physical implementation and the management of service providers, customers, and infrastructure with a security perspective.
Regarding security, the authentication of the users without revealing their real identity and exposing the location through multiple runs of dynamic charging is still an open discussion.

\subsection{Identity-Based Public Key Infrastructure}\label{subsec:ibe}
In this work, we use Identity-Based Public Key cryptography to secure the communication between the \acrshort{obu} and another entity that we describe in Section \ref{sec:models}, the \acrfull{csp}. 
In 1984, Shamir proposed the first idea of an Identity-Based scheme~\cite{shamir_ibe}, and in 2001 Boneh and Franklin introduced a functional \acrshort{ibe} based on Weil Rings on Elliptic Curves~\cite{boneh_ibe}. 
This public cryptographic system uses arbitrary public strings to encrypt messages, such as the receiver's email.
This simplifies the certificate management and the recipient only needs to obtain its private key from the \acrfull{pkg}, as it would do with a \acrfull{ca}. 
The \acrshort{pkg} is the trusted authority that publishes all the cryptographic tools and knows all the users' private keys. 

\subsection{Post-Quantum Cryptographic System}\label{subsec:pq_math}
The advent of \acrshort{pq} computing challenges the security assumptions of classic public key encryption. 
The necessity of a \acrshort{pq} encryption algorithm is more important than ever before quantum computing can threaten the privacy and security of the people. 
This is especially valid for resource-constrained devices with limited computational capabilities and batteries, which can cut off deploying a \acrshort{pq} algorithm in these instruments if not carefully designed. 
About the \acrshort{ibe} scheme, Ducas et al. presented an efficient scheme based on NTRU Lattices and \acrshort{rlwe} problem~\cite{ducas}. 
The authors show the possibility of deploying this system with coupled encryption and signature schemes. 
Here we briefly describe the mathematical tools that we exploit in our work.
We refer the reader to~\cite{ducas} for a deeper understanding and details.

\paragraph{NTRU Lattices} 
Before defining NTRU lattices, we need to define the following ring, where they are usually implemented:
\begin{equation}
\mathcal{R}\triangleq \mathbb{Z}_q[x]/(x^N+1),
\end{equation}
where $N$ is a power of 2. In this ring, we can define NTRU lattices as follows.
\begin{definition}[NTRU lattices]
Denote as $N$ an integer power of two, $q$ as a positive integer, and polynomials $f,g  \in \mathcal{R}$. Let $h = g*f^{-1} mod q$. The NTRU lattice associated to $h$ and $q$ is 
\[\Lambda_{h,q}= \{(u,v)\in\mathcal{R}^2 | u + v * h = 0 \mod q\}.\]
\end{definition}
$\Lambda_{h,q}$ is a full-rank lattice, generated by the rows of 
\[A=
\begin{pmatrix}
-\mathcal{A}_N(h) & \mathcal{I}_N \\
q\mathcal{I}_N & \mathcal{O}_N 
\end{pmatrix},\]
where $\mathcal{A}_N(h)$ is the N-dimensional anticirculant matrix, $q$ is the modulus, $\mathcal{I}_N$ and $\mathcal{O}_N$ are the Identity and Zero matrix, respectively. To perform standard lattice operations with better performance, we can efficiently found a basis computing $F,G \in \mathcal{R}$ such that:
\begin{equation}
    f*G-g*F = q.
\end{equation}
In this setting, the Hardness Assumption is derived by the indistinguishability of the quotient $f/g$ (two random polynomials $f,g \in \mathcal{R}$) from a random value in $\mathcal{R}_q$. 
Furthermore, in this scheme, we also use the \acrshort{rlwe} assumption: the distribution of $(h_i, h_is+e_i)$, with $h_i$ randomly chosen in $\mathcal{R}_q$, and $s$, $e_i$ as small polynomials, is indistinguishable from the uniform distribution.

\paragraph{\acrshort{ibe} operations} We describe the basic four operations necessary for the correct deployment of \acrshort{ibe} cryptographic system.
\begin{enumerate}
    \item MasterKeyGen: generates the Master Secret Key $B$ and the Master Public Key $h$. $B$ is a short basis of lattice $\Lambda_{h,q}$, making it a trapdoor for sampling short elements. The secret key is the prerogative of the trusted entity in the system.
    \item Extract: using the Master Secret Key, the trusted authority derives the secret keys for the users starting from their identity. The user secret key is composed of two small polynomials $s_1$ and $s_2$ such that $s_1 + s_2 * h = t$, where $t$ is the result of a hash function mapping the identity into the ring~$\mathcal{R}$.
    \item Encryption: this operation uses the public identity of a user as a public key to encrypt the message. Following the \acrshort{rlwe} assumption, the user chooses polynomials $r$, $e_1$, and $e_2$ with small coefficients and the ciphertext is
    \[(u = r*h + e_1, v = r*t + e_2 + \lfloor q/2\rfloor m).\]
    \item Decryption: using the private key assigned by the trusted entity, a user can decrypt a message received and encrypted with its public key (the identity itself). Starting from the ciphertext, the intended receiver can extract the plaintext as
    \[v-us_2 = rs_1 + e_2 + \lfloor q/2 \rfloor m - s_2e_1.\]
\end{enumerate}

In this protocol, we also use the digital signature scheme described in~\cite{ducas}, in which the key generation follows the aforementioned process. At the same time, the signing entity performs the Extract algorithm to perform the signature without the need for encryption. The verification step checks the norm of $(s_1, s_2)$ and the equality $s_1+s_2*h=H(message)$.

\section{System and Attack Model}\label{sec:models}
We present the models adopted in our work. In Section~\ref{subsec:network_model}, we describe the system and network model; whereas in Section~\ref{subsec:attack_model}, we present the adversary model.

\begin{figure*}[t]
  \centering
  \includegraphics[width=0.68\textwidth]{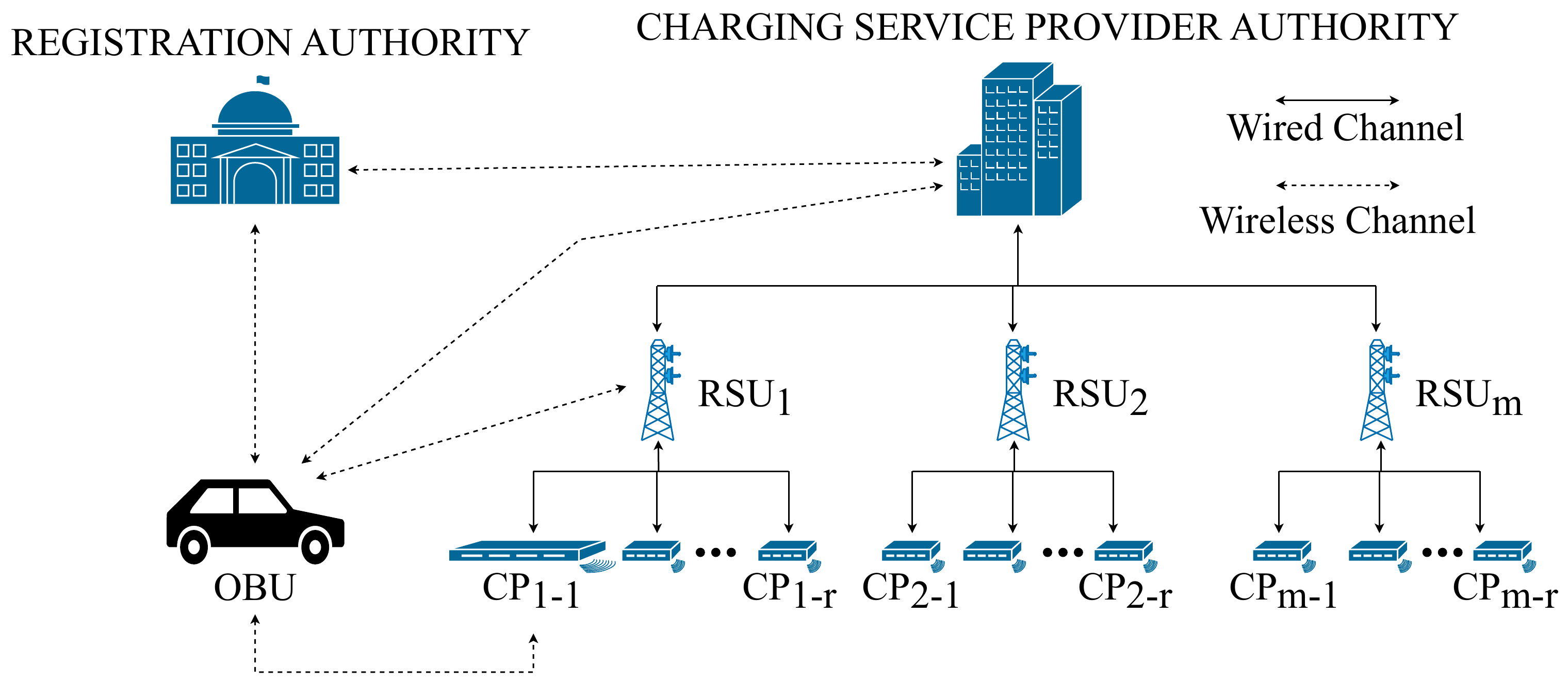}
  \caption{System model presenting the type of communication channel between the entities.}
  \label{fig:network_model}
\end{figure*}

\subsection{System Model}\label{subsec:network_model}
In this work, we employ a system model that emulates the toll station, so that it can deal with the different hardware capabilities in a post-quantum environment. The difference from the regular \acrshort{dwpt} model structure is the adoption of an entry point inspired by the roads with a toll. We depict the model in Figure~\ref{fig:network_model}. \acrfull{ra} is the only trusted entity in this system. It is in charge of guaranteeing the correct implementation and deployment of cryptographic keys. During the first steps of the protocol (registration phase), we consider secure communication between \acrshort{ra} and the other entities. The second main body is the \acrfull{cspa}. It comprehends a hierarchical structure, and it is in control of different \acp{rsu}, that in turn are in charge of a segment of \acp{cp}. These \acp{cp} are the most constrained part of the network and contain the hardware for the wireless power exchange with the vehicle's coil. Unlike other works, we consider a first pad dedicated to network operations involving the protocol's initial part. This is analogous to the toll roads where customers must first pass through a gate. The communication between components in the \acrshort{cspa} network is wired. Lastly, there is \acrshort{obu}, part of the \acrshort{ev}. It communicates with the elements of \acrshort{cspa} through wireless channels. It employs 5G network communication, while during an exchange with \acp{cp} it uses a \acrfull{dsrc} channel.

\subsection{Attack Model}\label{subsec:attack_model}
We consider the Dolev-Yao~\cite{dolevYao} adversary model,  well-known for public key protocols security assessments. In this model, an attacker can perform different actions like intercept, replay, compose, and forge messages. Without the cryptographic keys, they cannot decipher the ciphertext into plaintext. We assume the adversary wants to perform the two attacks peculiar to the \acrshort{dwpt} system: free-riding, and double-spending. 
In addition, the attacker tries to track the customer across different charging sessions. Thus, our scheme must be secure against attacks and protect the customers' privacy. 
The attacker can sniff the wireless network and tap the wired connection in the \acrshort{cspa} infrastructure. This last channel uses encryption to protect against eavesdropping~\cite{babuPUF} and \acrshort{cspa} sets it up before availability to the customers. Consequently, an attacker can't decipher the messages. In this sense, the wired network, in the scope of this work, is considered secure. On the opposite, the communication between \acrshort{obu}, \acrshort{cspa}, \acrshort{rsu}, and \acrshort{cp}s is vulnerable to attacks such as injection. The authentication protocol wants to secure it against wireless attacks in a \acrshort{pq} manner.

\section{Our Proposed Solution: DynamiQS}\label{sec:protocol}
\begin{figure*}[t]
  \centering
  \includegraphics[width=0.8\linewidth]{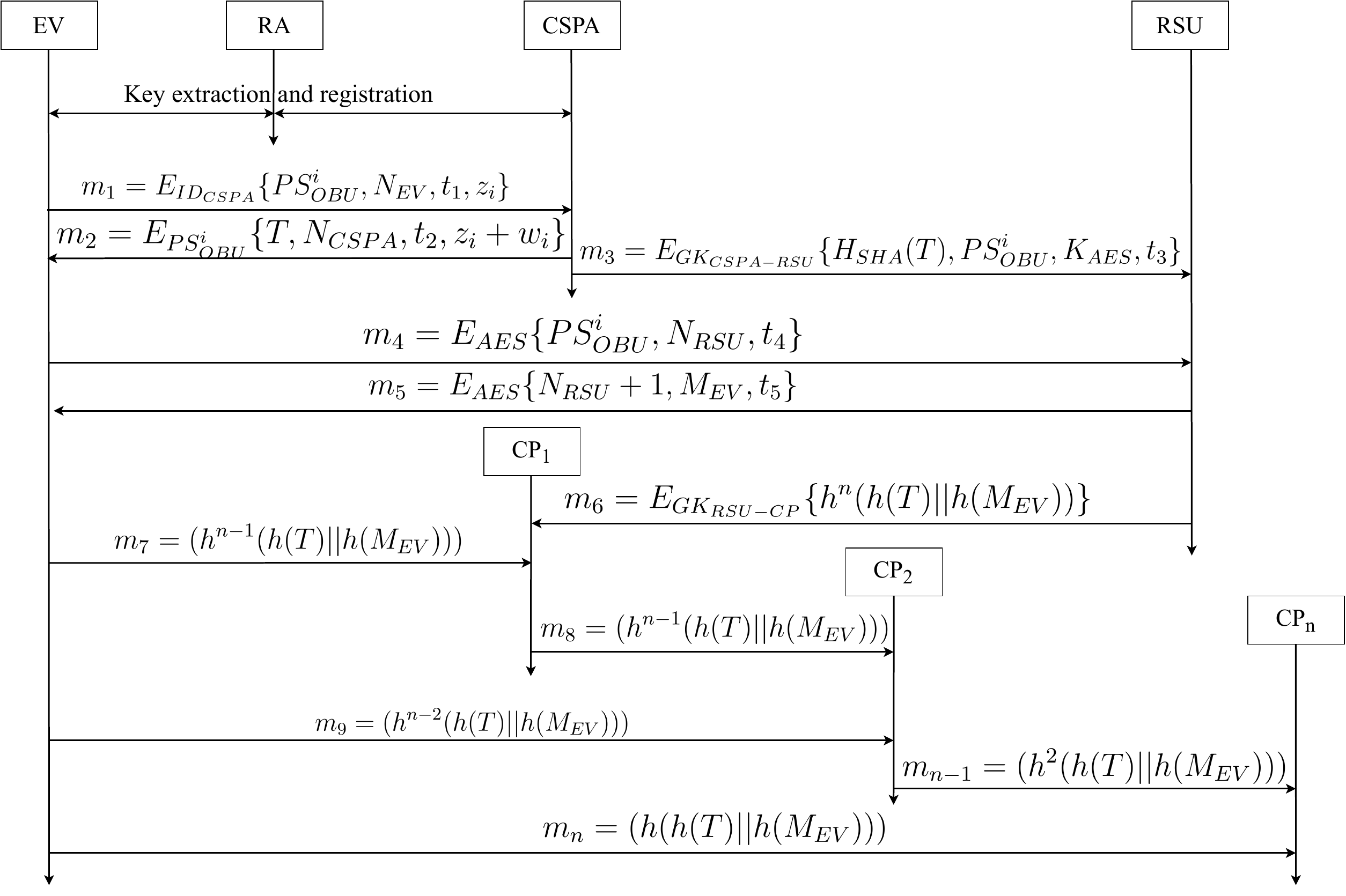}
  \caption{Flow chart of the authentication scheme between the \acrshort{ev} and the different parts of the \acrshort{cspa}'s network.}
  \label{fig:protocol}
\end{figure*}

In this section, we describe our protocol implementation, starting from Section~\ref{subsec:protocol-reg}, in which we describe the set-up and initialization phase. In Section~\ref{subsec:protocol-auth}, we present the authentication steps between \acrshort{obu}, \acrshort{cspa}, \acrshort{rsu}, and \acrshort{cp}s. Figure~\ref{fig:protocol} contains a graphical representation of a protocol run, while in Table~\ref{tab:symbols} we report the symbols used in this work.

\begin{table}[b]
  \caption{Symbols of the authentication protocol.}
  \label{tab:symbols}
  \begin{tabular}{cl}
    \toprule
    Symbol & Value\\
    \midrule
    $PS^i_{OBU}$ & Pseudonym for the $ith$ charging session \\
    $\mathbf{B}$ & Master Secret Key \\
    $h$ & Master Public Key\\
    $H$ & Hash function for Secret key generation \\
    $H_{SHA}$ & Hash function intended as SHA256 in this work\\
    $N,M$ & Nonce \\
    $T$ & Token \\
    $d_{EV}$ & Seed \\
    $a_i$ & Random number \\
    $K_{AES}$ & Session key for symmetric encryption with AES \\
    $SK_{entity}$ & Secret key for that entity \\
    $E_{ID}\{\}$ & Encryption with public ID \\
    $E_{GK_{E_1-E_2}}\{\}$ & Encryption with group key between the two entities \\
    $E_{AES}\{\}$ & Encryption with AES \\
    $t$ & Timestamp \\
  \bottomrule
\end{tabular}
\end{table}

\subsection{Set-up and Registration Phase}\label{subsec:protocol-reg}
Before providing authentication and charging capabilities, the \acrshort{ra} deploys and generates cryptographic parameters for the system and the users that want to register to the service provided by the \acrshort{cspa}. These phases happen offline, in a secure and trusted environment. The following steps describe the set-up and registration phases.

\paragraph{Generation of the Master Keys for \acrshort{ra}}
\acrshort{ra} handles the generation of the Master Secret Key and Master Public Key. It requires only two parameters: the polynomial degree $N$ as a power of 2, and the prime number $q$, congruent to $1 mod(2N)$. Afterward, \acrshort{ra} publishes the Master Public Key, defined as
\[
h = g * f^{-1}modq, 
\]
with parameters defined as in Section~\ref{subsec:pq_math}. Customers and \acrshort{cspa} use the Master Public Key during the encryption and decryption operations. Also, \acrshort{ra} stores securely the Master Secret Key
\[
\mathbf{B}=
\begin{pmatrix}
\mathcal{A}(g) & -\mathcal{A}(f) \\
\mathcal{A}(G) & \mathcal{A}(F) 
\end{pmatrix},
\] 
that is necessary for the registrants secret key generation. $\mathbf{B}$ is a short basis for the lattice $\Lambda_{h,q}$, and works as a trapdoor for sampling short elements $(s_1, s_2)$ without leaking information.

\paragraph{Extract the keys for registration} 
In this step, \acrshort{ra} generates the secret key for the customer. The user needs to send their identity. In the case of \acrshort{cspa}, the identity corresponds to the actual $ID_{CSPA}$. In the case of \acrshort{ev}, we generate the pseudonyms to be used during charging processes to provide anonymity. The pseudonym generation starts with \acrshort{ra} choosing a seed and a random number, $sd_{EV}$ and $a_i$, respectively, and generate the pseudonym
\[PS^i_{OBU} = H_{SHA256}(ID_{EV} || d_{EV} * a_i).\]
The \acrshort{obu} only needs the parameters $sd_{EV}$ and $a_i$ to compute the pseudonyms on its own when needed. \acrshort{ra} will then compute the private key through the \textit{extract} phase for different pseudonyms and send them to the \acrshort{ev}. The \textit{extraction} takes as input the Master secret key and the identity, or pseudonym, of the registrant. Additionally, it uses a hash function defined as $H: {0,1}^* \rightarrow \mathcal{Z}^N_q$. The output is a tuple of small polynomials $(s_1, s_2)$ as we describe in Section~\ref{subsec:pq_math}, and the private key $SK_{ID}$ is $s_2$. We denote as $SK_{CSPA}$ and $SK_{PS^i_{OBU}}$ the private keys of \acrshort{cspa} and private key of \acrshort{ev}'s pseudonym for the $ith$ protocol run, respectively. Additionally, \acrshort{ra} generates a couple of random numbers $(z_i, w_i)$ associated with each pseudonym. These are essential for the first authentication step between \acrshort{ev} and \acrshort{cspa}. In fact, \acrshort{cspa} receives from \acrshort{ra} a copy of the dataset containing the pseudonyms of all the vehicles registered and the random numbers $(z_i, a_i)$. In this way, there is a common secret between \acrshort{cspa} and the vehicles without revealing information about the identity. In Section~\ref{sec:performance}, we analyse the storage performances.

\subsection{Authentication Phase}\label{subsec:protocol-auth}
When the customer wants to charge their vehicle, they need to pass through the first pad, which is dedicated to the first cryptographic exchange. The authentication starts from the first level, \acrshort{cspa}, down to the first \acrshort{cp}, passing to \acrshort{rsu}. We describe the message exchange for the proposed authentication protocol, considering the $ith$ run and use of the corresponding pseudonym by \acrshort{ev}. 
\begin{enumerate}
    \item The process begins with \acrshort{ev} sending the first message to \acrshort{cspa} containing the pseudonym, a nonce for the AES session key generation, timestamp, and the common secret via its \acrshort{obu}. The message is encrypted with the \acrshort{cspa}'s identity, and thanks to the \acrshort{ibe} scheme, only the intended receiver with the correct private secret key can decrypt it.
    \begin{equation}
        \acrshort{ev} \rightarrow \acrshort{cspa}:
        m_1 = E_{ID_{CSPA}}\{PS^i_{OBU}, N_{EV}, t_1, z_i\}. 
    \end{equation}

    \item \acrshort{cspa} decrypts the cipher-text and gets the plain-text. It stores the pseudonym and the value $N_{EV}$. Having the pseudonym allows \acrshort{cspa} to check for the possibility of reuse by a malicious user or an attacker. \acrshort{cspa} saves the extracted $PS^i_{OBU}$ in a lookup table for revocation and finds the corresponding value $(z_i, w_i)$. After that, \acrshort{cspa} creates a token $T$ for the last phase of the authentication between the \acrshort{ev} and \acrshort{cp}s. Additionally, \acrshort{cspa} generates a nonce $N_{CSPA}$ for the session key derivation. Finally, it sends these two values to \acrshort{ev}, encrypting the message with the pseudonym as identity. The session key derivation is a SHA hash of the concatenation of the two nonces as $K_{AES}=H_{SHA}(N_{EV} || N_{CSPA})$.
    \begin{equation}
        m_2 = E_{PS^i_{OBU}}\{T, N_{CSPA}, t_2, z_i+w_i\}.
    \end{equation}
    After receiving the message, \acrshort{ev} can verify the operation on $z_i$, generate the session key and get the token's hash. Here, the first layer of authentication between \acrshort{cspa} and \acrshort{ev} is completed. 

    \item At the same time, \acrshort{cspa} sends the SHA hash of the token to the first \acrshort{rsu}, along with the pseudonym and the session key. \acrshort{cspa} sends this message in the wired channel, encrypting it with the group key established between \acrshort{rsu}s.
    \begin{equation}
        m_3 = E_{GK_{CSPA-RSU}}\{H_{SHA}(T), PS^i_{OBU}, K_{AES}, t_3\}.
    \end{equation}
    The \acrshort{rsu} extracts the values and waits for a message from \acrshort{ev} to continue the process.

    \item Next, \acrshort{ev} encrypts the pseudonym and another nonce to use with \acrshort{rsu} as a proof of work. It uses AES symmetric cipher using the session key previously computed. 
    \begin{equation}
        m_4 = E_{AES}\{PS^i_{OBU}, N_{RSU}, t_4\}.
    \end{equation}

    \item \acrshort{rsu} needs the correct session key to decrypt the message $m_4$ and get the parameters inside. The nonce $N_{RSU}$ proves its decryption ability, sending it back augmented by 1. Moreover, \acrshort{rsu} computes a nonce, $M_{EV}$, necessary for the hash chain generation, that is \[H^n_{SHA}(H_{SHA}(T)||H_{SHA}(M_{EV})),\]
    with $n$ a large enough number to cover all the pads during the authentication. The value $n$ is the number of times that the parameters are hashed. Finally, \acrshort{rsu} sends message $m_5$ to the \acrshort{ev}:
    \begin{equation}
        m_5 = E_{AES}\{N_{RSU}+1, M_{EV}, t_5\}.
    \end{equation}

    \item Promptly, the\acrshort{rsu} also sends the head of the hash chain to the first \acrshort{cp}, encrypted with the group key between the two entities. At this point, the authentication between \acrshort{ev} and \acrshort{rsu} is concluded.
    \begin{equation}
        m_6 = E_{GK_{RSU-CP}}\{ h^{n}(h(T) || h(M_{EV}))\}.
    \end{equation}

    \item The last part involves the use of the hash chain as a means of authentication, using the value immediately before the last one used. In this case, \acrshort{ev} sends to the first encountered pad the value:
    \begin{equation}
        m_7 = (h^{n-1}(h(T) || h(M_{EV}))).
    \end{equation}
    The \acrshort{cp} checks the received value against the one got by \acrshort{rsu}, in turn hashing $m_7$:
    \[ H_{SHA}(H^{n-1}_{SHA}(H_{SHA}(T)||H_{SHA}(M_{EV})) \]
    \[ == \] 
    \[ H^{n}_{SHA}(H_{SHA}(T)||H_{SHA}(M_{EV}),\]
    where the symbol "==" represents the equality test. If values correspond, the \acrshort{cp} charges the vehicle. For completeness, we report the message exchange with the second \acrshort{cp} since consecutive pads have the same behavior. The first \acrshort{cp} sends the next known value of the chain, so the same as the one received by \acrshort{ev}:
    \begin{equation}
        m_8 = (h^{n-1}(h(T) || h(M_{EV})))
    \end{equation}
    to the second pad while the vehicle sends the previous hash in the chain:
    \begin{equation}
        m_9 = (h^{n-2}(h(T) || h(M_{EV}))).
    \end{equation}
\end{enumerate}
At this point, the authentication and the power exchange continue until the charging lane ends:
\begin{equation}
    m_{n-1} = (h^{2}(h(T) || h(M_{EV}))),
\end{equation}
\begin{equation}
    m_n = (h(h(T) || h(M_{EV}))).
\end{equation}

\section{Security Analysis}\label{sec:security}
In Section \ref{subsec:informal} we informally describe the protocol's security against the post-quantum threats. In Section \ref{subsec:formal}, we provide a formal security analysis using BAN logic~\cite{ban} and Scyther verification tool~\cite{scyther} for authentication schemes.

\subsection{Informal Analysis}\label{subsec:informal}
The \acrshort{pq} resistance of the protocol is inherited from the \acrshort{ibe} based on lattices. The classic \acrshort{ibe} scheme is vulnerable to the future \acrshort{pq} machines that can break the encryption and key derivation operations~\cite{verchyk}. Also, the symmetric encryption based on AES, and the hash chain are \acrshort{pq} secure. Following, we informally present the protocol's security against the particular threats of this scenario.
\paragraph{Free-riding} The vehicle asking for a charge must also authenticate at pad level, and the hash chain value is valid only for the next expected car. An attacker can't guess the current hash in the chain of another user. Also, the pads don't accept an already-used hash value. In this way, it is impossible to get a free charge.
\paragraph{Double spending} In the same way, the pads expect a specific hash value that the car can use only once, so it is not allowed for the customer to send an old message to get a charge for free.
\paragraph{Identity spoofing} The use of pseudonyms generated in advance and in a secure environment, stored in the trusted platform in the \acrshort{obu}, makes spoofing and using the real identity of a customer impossible. The real identity is never revealed during the authentication steps.
\paragraph{Location tracking} The use of the same identifier over multiple charging sessions could jeopardize the user's privacy, allowing \acrshort{cspa} to determine the path and the car's location. In this matter, using a different pseudonym for each charging lane and session makes this tracking ineffective. Furthermore, linking different pseudonyms and relating them to the same owner is impossible.
\paragraph{Message spoofing} An attacker can try to impersonate another entity, but without the correct private key to decrypt a message, it is impossible to masquerade the identity. Only the registered cars and \acrshort{cspa} have the shared secrets to start a conversation and the possibility to know the messages' content.
\paragraph{Forward secrecy} A vehicle exchanges messages using pseudonyms and ephemeral session keys for each protocol run. If an attacker gets access to the private key of a pseudonym, it can get access only to the message of that exchange, also without being able to generate the session key. Conversely, if the attacker obtains a session key, it does not harm future communications. 

\subsection{Formal Analysis}\label{subsec:formal}
\paragraph{BAN Logic}
BAN Logic is a formal procedure to prove resistance against replay attacks and identify possible flaws in the protocol~\cite{ban}. It uses formal algebra to describe the authentication steps through the means of \textit{freshness} and \textit{shared secret}. The first principle states that when a message with a nonce is sent for the first time, and a response based on that nonce is received, the last message is fresher than the previous one. The \textit{shared secret} recalls that if a message encrypts a valid secret between two trusted entities, this message must have been sent by one of these two parties. \\
From the analysis perspective, from the protocol's initial state, the relationships between stages or messages are determined by the \textit{belief} of the freshness and secrecy of the content. At the final step, it is possible to determine improvements or state the formal proof of security. Table~\ref{table:ban_constructs} reports the BAN constructs used in the following formal analysis.

\begin{table}[b!]
\begin{center}
    \caption{BAN Constructs and Rules used in the formal analysis.}
    \begin{tabular}{cp{4.5cm}}
        \hline
        \textbf{Notation} & \textbf{Description} \\
        \hline
        $P |= X$ & P believes X, so P thinks that X is true. \\
        \hline
        $P <| X$ & P sees message X. \\
        \hline
        $\{X\}K$ & encrypted with key K.  \\
        \hline
        $P |\sim X$ & P once said X. \\
        \hline
        $\#(X)$ & X is fresh. \\
        \hline
        $P <- K -> Q$ & P and Q shared a secret key. \\
        \hline
        $P=X=Q$ & X is a secret known only by P and Q (or other trusted parties) \\
        \hline
        Shared Key Rule & $\frac{P |= Q <- K -> P, P <| \{x\}K}{P |= Q |\sim X}$, If P believes that K is a good K, and P sees X encrypted with K, then P believes that Q once said X.  \\
        \hline
        Nonce Verification Rule & $\frac{P |= \#(X), P |= Q |\sim X}{P |= Q |= X}$, the only formula to promote $|\sim$ to $|=$, says that P believes X to be recent, and Q said X, then P believes that Q believes X. \\
        \hline
        Freshness Rule & $\frac{P |= \#(X)}{P |= \#(X, Y)}$, if part of the formula is fresh, the entire formula is believed to be fresh. \\
        \hline
    \end{tabular}
    \label{table:ban_constructs}
\end{center}
\end{table}

Our scheme starts with an encrypted message, making the symmetric key exchange and derivation much more secure. Following the Freshness rule, we state:
\begin{equation}\label{eq:ban1}
CSPA <| \{m_1\}K_{IBE_{CSPA}}, \frac{CSPA |= \#(t_1)}{CSPA |= \#(N_{EV}, t_1)}.
\end{equation} 
Then, by the Nonce Verification rule,
\begin{equation}\label{eq:ban2}
\frac{CSPA|= \#(N_{EV}), CSPA|=EV|\sim N_{EV}}{CSPA|=EV|=N_{EV}}.
\end{equation} 
In Equation \ref{eq:ban1}, we define the message $m_1$ as encrypted with the public key of \acrshort{cspa}, which recognizes the timestamp's freshness. As the Freshness rule implies, the nonce $N_{EV}$ is fresh, and the Nonce Verification (Equation \ref{eq:ban2}) rule states the security of it. 
On the opposite, the same happens on EV's side:
\begin{equation}
EV <| \{m_2\}K_{IBE_{EV}}, \frac{EV|= \#(t_2)}{EV |= \#(N_{CSPA}, t_2)},
\end{equation}
and consequently $N_{CSPA}$ is a good nonce for the authentication,
\begin{equation}
\frac{EV|= \#(N_{CSPA}), EV|=CSPA|\sim N_{CSPA}}{EV|=CSPA|=N_{CSPA}}.
\end{equation}
The following essential messages for the authentication verification are $m_4$ and $m_5$, for which we can apply the Shared Key Rule (Equation \ref{eq:ban3}). In this case, \acrshort{rsu} and \acrshort{ev} exchange encrypted messages with the AES key. \acrshort{rsu} can verify the knowledge of the key by \acrshort{ev} simply by decrypting $m_4$, considering the key as a secure shared secret. The previous steps already prove the security of the AES key. Consequently:
\begin{equation}
RSU <- K_{AES} -> EV, RSU <| {m_4}K_{AES}, RSU|=\#(t_4), 
\end{equation}
\begin{equation}\label{eq:ban3}
\frac{RSU|=RSU<-K_{AES}->EV, RSU<|{m_4}K_{AES}}{RSU|=EV|\sim m_4}.
\end{equation}
Following the Freshness rule and Nonce Verification rule, applied as before, they prove the security of the message. The same applies for $m_5$, so \acrshort{ev} can verify the authenticity of \acrshort{rsu}, which correctly decrypts the message and sends back the update nonce encrypted with the correct key. The message exchange with the \acrshort{cp}s doesn't require proof of security and not repeatability due to the use of the hash-chain. Furthermore, only the intended user can generate and send the correct hash value.

\begin{figure}[!t]
  \centering
  \includegraphics[width=0.8\linewidth]{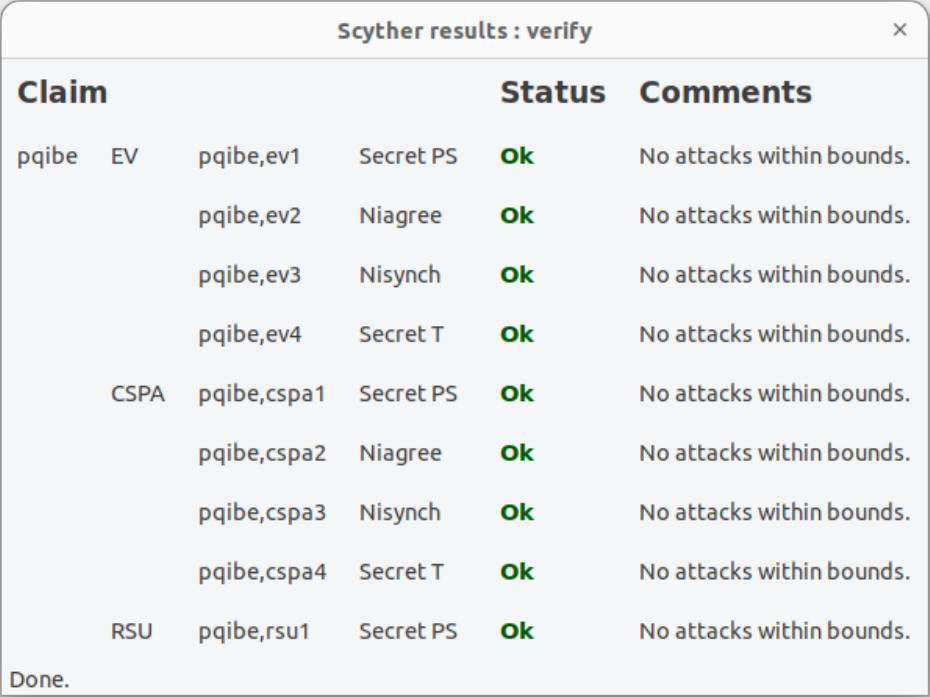}
  \caption{Scyther results for our protocol.}
  \label{fig:scyther}
\end{figure}

\paragraph{Scyther results}
We investigate the protocol's security using the Scyther tool~\cite{scyther} and document the code in the GitHub project, indicated in the next Section. As Figure~\ref{fig:scyther} shows, the tool confirms the security and secrecy of the pseudonym during a protocol run. The \acrshort{ra} attests the accredited entities subscribed to the \acrshort{ibe} service. In this scheme, the identity is the public key, so it is impossible to pose as another entity.

\section{Performance Analysis}\label{sec:performance}
We describe the protocol in terms of computation and communication performance in Section~\ref{subsec:computation} and Section~\ref{subsec:communication}, respectively. In Section~\ref{subsec:experiment}, we report the results of the protocol simulation and analysis for computation and communication costs.

Being the first work on \acrfull{pq}, a comparison with state-of-the-art protocols is challenging. This is due to the different model we use and the cryptographic operations to perform, which are different from any other related work for the novel application of \acrshort{pq} cryptographic tools on authentication for dynamic wireless EV charging. In this matter, we provide an analysis of the authentication scheme that can work on constrained devices minding the \acrshort{pq} advent.

\begin{table}[b]
    \begin{center}
        \caption{Primitive time required on Cortex M-0.}
        \begin{tabular}{cp{1.5cm}p{1.5cm}p{1.5cm}}
            \hline
            \textbf{Primitive} & \textbf{Symbol} & \textbf{Cycles} & \textbf{Time (ms)} \\
            \hline
            \acrshort{ibe} encryption & $t_{IBE_{enc}}$ & 3 297 380 & 103.00 \\
            \hline
            \acrshort{ibe} decryption & $t_{IBE_{dec}}$ & 1 155 000 & 36.00 \\
            \hline
            AES-256 operation & $t_{AES-256}$ & 10 611 & 0.33 \\
            \hline
            SHA-256 operation & $t_{SHA-256}$ & 11 561 & 0.36 \\
            \hline
        \end{tabular}
        \label{tab:computation}
    \end{center}
\end{table}

\begin{table*}
    \begin{center}
        \caption{Computation and communication theoretical cost for each message, considering the time to send it through the network. The last two rows represent the total cost for the first pad and the asymptotic form.}
        \begin{tabular}{cp{2cm}p{5cm}p{3.5cm}p{3cm}}
            \hline
            \textbf{Message} & \textbf{Channel} & \textbf{Computation Cost ($ms$)} & \textbf{Communication Cost ($B$)} & \textbf{Sending Time ($\mu s$)} \\
            \hline
            $m_1$ & 5G & $t_{IBE_{enc}}+t_{IBE_{dec}}+t_{SHA-256}=139.36$ & 128 & 10.24 \\
            $m_2$ & 5G & $t_{IBE_{enc}}+t_{IBE_{dec}}+2t_{SHA-256}=139.36$ & 128 & 10.24 \\
            $m_3$ & Fast Ethernet & $t_{AES-256}+t_{SHA-256}=0.69$ & 128 & 10.24 \\
            $m_4$ & 5G & $t_{AES-256}=0.33$ & 96 & 7.68 \\
            $m_5$ & 5G & $t_{AES-256}=0.33$ & 96 & 7.68 \\
            $m_6$ & Fast Ethernet & $t_{AES-256}+t_{SHA-256}=0.69$ & 32 & 2.56 \\
            $m_7$ & DSRC & $n\times t_{SHA-256}=n\times 0.36$ & 32 & 9.48 \\
            $m_8$ & DSRC & Forwarding and no computation & 32 & 9.48 \\
            $m_9$ & DSRC & $t_{SHA-256}=0.36$ & 32 & 9.48 \\
            \hline
            Total $CP_1$ & --- & $281.12 + n\times 0.36$  & 640 & 58.12 \\
            \hline
            Total Asymptotic & --- & $280.76 + n \times 0.36 + T_{CP}$ & $576 + n\times 64$ & $48.64+n\times 18.96$ \\
            \hline
        \end{tabular}
        \label{tab:cost_summary}
    \end{center}
\end{table*}

\subsection{Computational Performance}\label{subsec:computation}
Our performance calculation is based on the work of G{\"u}neysu et al.~\cite{guneysu}. The authors demonstrate the practicability of \acrshort{ibe} scheme in embedded devices with low-cost hardware. Their experiments show that \acrshort{ibe} scheme for post-quantum can efficiently provide a security level of 80 bits, requiring only 103 ms for the encryption and 36 ms in the decryption phase. We report here the computation and results for the Cortex M-0 hardware. In addition, to keep consistency with the calculation performed on real hardware by~\cite{guneysu}, we also use the results obtained by~\cite{cifra}, who computed the time of operations such as AES and hashing on the same type of hardware. The results on the Cortex M-0 are the worst case, and~\cite{guneysu} provides results for Cortex M-4 with much better performance. We want to consider the time taken by the Cortex M-0 with a 32 MHz clock and cheap hardware. This time represents the slower implementation and worst-case scenario. Table~\ref{tab:computation} reports the cycles (31.25 ms per cycle) required for the operations included in our protocol: encryption and decryption with \acrshort{pq}-\acrshort{ibe}, symmetric operations on AES-256, and hashing with SHA-256. The total time required by the protocol considers the encryption and decryption with \acrshort{pq}-\acrshort{ibe} of the two first messages, the encryption and decryption with a symmetric scheme of $m_3$, $m_4$, $m_5$, and $m_6$. Furthermore, the total number of hash operations before the hash chain is 5. Here, we provide two different results: only the first pad with the hash chain computation and the asymptotic result considering the entire \acrshort{cp} lane. In the first case, we have a total time of $281.12$ ms and the time $n \times T_{SHA-256}$, with $n$ the number of pads, to build the hash chain by \acrshort{ev} and \acrshort{rsu} (in a simultaneous process, also considering the higher \acrshort{rsu}'s computation capabilities). This time considers already the check of the hash sent by \acrshort{ev} to the fist \acrshort{cp}. As an example, we can have a lane of 100 pads. This would take $T_{SHA-256}\times n$ time, equal to $36.1$ ms, for the hash chain computation. The chain time is added to the first pad computation for a total of $317.12$ ms. In the second case, we consider the time taken up to the message $m_6$, while the authentication phase between \acrshort{ev} and \acrshort{cp}s is defined in an asymptotic way as follows:
\begin{equation}
    T_{CP}=\frac{(n^2+n)}{2}\times T_{SHA-256}.
\end{equation}
The overall time is $280.76 ms + n \times T_{SHA-256} + T_{CP}$. In Table~\ref{tab:cost_summary}, we delineate the time for each message and the two different cases. \par
Concerning the storage capacity of the devices involved, and the values needed in the first authentication layer between \acrshort{ev} and \acrshort{cspa}, we compute the storage occupation of the pseudonyms and the shared secrets. This analysis supports the possibility of implementing this scheme in real hardware. We can consider the pseudonyms and the value $(z_i, a_i)$ as 32 Bytes and 64 Bytes being hash digest and numbers, respectively. Moreover, we consider \acrshort{cspa} as an entity with 10 million cars registered to its service. Each vehicle has 3650 associated pseudonyms, relative private keys, and secrets. This number is largely enough, being a charge permission for every day for ten years of the vehicle's life. Having 96 Bytes for each pseudonym means 350.4 KBytes per vehicle, for around 3.5 TBytes of data managed by \acrshort{cspa}. A value that a large entity can easily handle. In the \acrshort{ev}'s system, it is the 350.4 KBytes additionally to the storage of the small polynomial $s_2$ of the secret key. A value that a constrained device on the car can address.

\subsection{Communication Performance}\label{subsec:communication}
The communication cost considers the messages' length and the technology's nominal bit rate. In Table~\ref{tab:channels}, we point out the channels and their capacity. As in Section~\ref{subsec:computation}, we divide the two cases considering only the first pad or an asymptotic lane. The length of Nonces, Digests, and AES block encryption is 32 Bytes, assuming hash and symmetric encryption at 256 bits. The same applies to timestamps, treated as an integer. The total communication cost of the first pad is $640$ Bytes, from message $m_1$ to $m_7$. In asymptotic form, we consider up to message $m_6$ a total amount of $576$ Bytes plus $n\times 64$, with $n$ number of pads in the lane. Table~\ref{tab:cost_summary} summarized the total computation and communication cost for each message, giving an overall overview of the theoretical cost of the protocol. \par
Thanks to the obtained results, we can give a theoretical formulation to compute the length of the first pad. Lastly, we consider the computational and communication time to estimate the length of the first pad, which functions as a gate for the authentication and enabler of the charging process. It must be long enough to permit the authentication steps that take place at the beginning. As expected, it depends on the maximum velocity we want the vehicle to reach and the number of pads. We define the length as:
\begin{equation}\label{f:pad_len}
    L_{pad}=v_{max}\times (281.12+n\times 0.36).
\end{equation}
These values are design choices, considering the trade-off between the time the authentication takes at the protocol's beginning, the pad's length, and the maximum velocity allowed. Higher velocity requires a longer first pad, but this length can decrease with fewer pads in the lane at the expense of the total charge exchanged. 


\begin{table}[t]
    \begin{center}
        \caption{Channel mediums and corresponding bitrate.}
        \begin{tabular}{cp{2cm}p{2cm}}
            \hline
            \textbf{Channel} & \textbf{Technology} & \textbf{Avg. Bitrate (Mbps)} \\
            \hline
            \acrshort{cspa}'s wired channel & Fast Ethernet & 100 \\
            \hline
            Generic wireless channel & 5G & 100 \\
            \hline
            \acrshort{cp}-\acrshort{ev} wireless channel & DSRC with 802.11p & 27 \\
            \hline
        \end{tabular}
        \label{tab:channels}
    \end{center}
\end{table}

\subsection{Experiments}\label{subsec:experiment}
\begin{figure}[t]
  \centering
  \includegraphics[width=\linewidth]{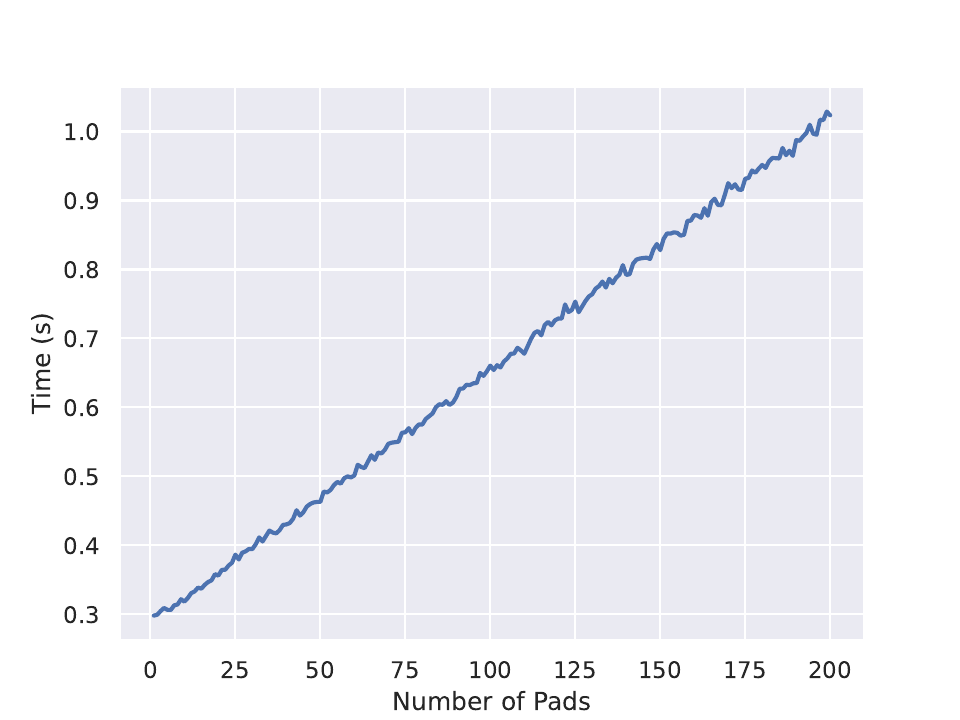}
  \caption{Time impact against the incremental number of pads in the charging lane.}
  \label{fig:pads_vs_speed}
\end{figure}

The theoretical analysis is supported by the experiments we perform using the network simulator \textit{ns-3.38}~\cite{ns3}. In the network simulator, we model the communication between the different entities using \textit{\acrfull{csma} Ethernet layer} for wired communications, \textit{PointToPoing communication} to simulate the 5G wireless channel with \acrshort{ev} and \acrshort{cspa}, and \acrshort{ev} and \acrshort{rsu}. The channel representing the \acrshort{dsrc} communication is simulated as \acrshort{csma} due to the limitation of the wireless channel provided by the ns-3.38 library; it can contain only up to 18 nodes, while we want the possibility to model a large number of \acrshort{cp}s. Furthermore, we use a delay in the communication corresponding to the computational time to mimic the cryptographic operations at a specific node. We study the impact of the number of pads on the total time taken by the protocol. Figure~\ref{fig:pads_vs_speed} shows the linear increments on the protocol time up to 200 \acrshort{cp}s. The time for the first \acrshort{cp} follows the slightly lower theoretical result of 281.12 plus the time for the chain construction. The simulated time is around 298.11 ms in ns-3.38. 

Assuming the large number of 200 pads, the total time grows to around one second. This time must be considered in the road scenario, where hovering over 200 pads in this interval is impossible if the customer wants to get the charge physically. Therefore, the computational and communication time can be achieved and used in a real scenario. \par
Additionally, we study the minimum length of the first pad depending on the vehicle's speed and the total number of \acrshort{cp}s. We perform this calculation in a script considering the formula~\ref{f:pad_len}. In Table~\ref{tab:length}, we show the length, in meters, against the maximum speed of the vehicle during the first communication exchange. Considering only the usual average speed of 130 Km/h on a highway, the length of the first charging pad varies from 10.20 to 12.79 meters. This is similar to the size of a toll booth, but where the driver usually must slow down to low speed. Furthermore, the subsequent pads' length is much smaller than the first \acrshort{cp} due to the low complexity and low computational time operations to perform. Finally, we provide the simulation code and analysis on GitHub\footnote{https://anonymous.4open.science/r/PQ-IBE-DWPT-Auth-31CF/README.md}. 

\begin{table}
    \begin{center}
        \caption{Length of the first pad depending on the vehicle's speed and total number of pads. The speed is measured in km/h, while the unit measure of the length is in meters.}
        \begin{tabular}{c|p{0.8cm}p{0.8cm}p{0.8cm}p{0.8cm}p{0.8cm}}
            \hline
            \diagbox{\textbf{Speed}}{\textbf{\# Pads}} & \textbf{10} & \textbf{50}  & \textbf{100}  & \textbf{150}  & \textbf{200} \\
            \hline
            \textbf{10} & 0.79 & 0.83 & 0.88 & 0.93 & 0.98 \\
            \hline
            \textbf{30} & 2.38 & 2.50 & 2.65 & 2.80 & 2.95 \\
            \hline
            \textbf{50} & 3.97 & 4.17 & 4.42 & 4.67 & 4.92 \\
            \hline
            \textbf{70} & 5.56 & 5.84 & 6.19 & 6.54 & 6.89 \\
            \hline
            \textbf{90} & 7.14 & 7.50 & 7.95 & 8.40 & 8.85 \\
            \hline
            \textbf{110} & 8.73 & 9.17 & 9.72 & 10.27 & 10.82 \\
            \hline
            \textbf{130} & 10.32 & 10.83 & 11.49 & 12.14 & 12.79 \\
            \hline
        \end{tabular}
        \label{tab:length}
    \end{center}
\end{table}

\section{Discussion}\label{sec:discussion}
Developing wireless charging infrastructure to support driving charging is still ongoing. To preserve user's privacy and security, researchers are trying to build authentication protocols and billing schemes that cover the real identity of the customers. Testing such schemes is still out of the possibility of researchers, who can only simulate the behavior and costs of their work. At the same time, the physical charging process is improving but still limiting a vehicle's possible speed and trajectory. In this sense, the charging process is the actual bound for a protocol. In general, the message exchange is faster than the power exchange, allowing more margin for the connectivity computational time. \par
Concerning this work, we show the possibility of using \acrlong{pq} \acrlong{ibe} without seriously impacting the process. However, this cryptographic tool's main limitations are the key size and the impossibility of using a signature scheme efficiently. The first problem can be bypassed with the key management at \acrshort{ra} side, with the key exchange and derivation during an offline phase. The second problem is more challenging to solve due to the mathematical complexity and impossibility of using it on constrained devices. Hence, mathematicians and cryptographers are working on further improving the \acrshort{pq} digital signature schemes so that it will be possible to implement in low-end devices. Finally, we don't provide a performance comparison with the state-of-the-art protocols due to the different cryptographic tools used. This is the first \acrlong{pq} authentication protocol proposed in the dynamic charging scenario for electric vehicles. We use complex mathematical tools and heavy operations. Instead, the state-of-the-art protocols don't provide security against the \acrshort{pq} threats and use faster and less involved computations, such as EX-ORing and hash functions. 

\begin{table*}[tp]
    \begin{center}
        \caption{Cryptographic primitives used in the most recent related works between \acrshort{ev} and \acrshort{cspa} network entities}
        \begin{tabular}{cp{6.9cm}p{6.9cm}}
            \hline
            \textbf{Work} & \textbf{Cryptographic primitives} & \textbf{Services} \\
            \hline
            Raveendra et al.~\cite{babuRobust} & Hash functions, XOR operation, ECC Decisional Diffie-Hellman & Mutual authentication \\
            \hline
            Raveendra et al.~\cite{babuPUF} & Hash functions, XOR operation, use of Physical Unclonable Functions & Mutual authentication, seamless handover, security against machine learning attacks \\
            \hline
            Raveendra et al.~\cite{babuHandover} & Hash functions, XOR operation & Mutual authentication, seamless handover \\
            \hline
            Asokraj et al.~\cite{asokraj} & \acrshort{ibe} public key, bilinear pairing, hash functions, symmetric session key & Mutual authentication, on-demand charging \\
            \hline
            DynamiQS & \textbf{\acrlong{pq} \acrshort{ibe} public key}, \acrshort{rlwe}, hash functions, symmetric session key & Mutual authentication, robustness against \acrlong{pq} computing \\
            \hline
        \end{tabular}
        \label{tab:cost_summary}
    \end{center}
\end{table*}

\section{Related Work}\label{sec:related}
This Section lists the main works done in the \acrshort{dwpt} scenario about authentication protocols. It is worth mentioning that, to the best of our knowledge, none of the schemes presented in literature so far is \acrshort{pq} resistant.

Back in 2017, Zhao et al.~\cite{free-riding} and Rabieh et al.~\cite{double-spending} proposed the two schemes against the free-riding attack and double-spending attack, respectively. The former uses public-key encryption and signing to secure and keep the customer's privacy, which can be charged only with the presentation of an authentication token request to the bank. Zhao et al. included a \acrfull{ra} in charge of the cryptographic parameters of their system. Additionally, they solved the free-riding threat by periodically checking the vehicle's battery level and certifying that it doesn't increase during a time interval in which the car is unauthenticated. Instead, the latter removed the use of \acrshort{ra} as a central authority and based their security on blind signatures and XOR operations. In their scheme, customers need to buy tickets for charging in advance. Also, they claim anonymity and privacy, solving the double-spending problem.  
Roman et al.~\cite{roman}, in 2020, proposed an authentication protocol in a cloud environment composed of a fog server, \acrfull{rsu}, and \acrshort{cp}s structured in a hierarchical architecture. This is the most common network model used in \acrshort{dwpt} research. Their scheme uses short signatures and blind signatures on bilinear pairing. As~\cite{double-spending}, users must buy tickets in advance to claim a charge.

Raveendra et al.~\cite{babuRobust} developed a robust and lightweight authentication protocol the last year. In this scheme, vehicles directly register to the \acrfull{ccs}, which initializes the system working on elliptic curve cryptography using signing. In this model, also \acrfull{fs}s need to register to the \acrshort{ccs}. \acrshort{fs}s are in charge of different \acrshort{rsu}, mutually authenticated with the cars when they need to charge. The messages are encrypted between \acrshort{ev}, \acrshort{fs}, and \acrshort{rsu}, while the authentication between \acrshort{ev} and \acrshort{cp}s happens through hash chains, as for the older protocols in the literature. The authors provide security proof using Scyther tool~\cite{scyther} and a MIRACL-based testbed experiment. Some of the authors of this work also proposed a protocol based on \acrfull{puf}~\cite{babuPUF}. In this way, the physical implementation of the hardware assures the entity's authenticity. The strong advantage is the resistance against physical attacks. 

Raveendra et al.~\cite{babuHandover} and Asokraj et al.~\cite{asokraj} proposed a new paradigm for seamless handover and on-demand charging requests for dynamic charging. The first work provides mutual authentication, and the authors claim computation and communication performance improvements. The second work allows the customer to start and stop the charging process at will, without needing to buy the amount of charge in advance. At the same time, the authors assert that this feature only weighs a negligible amount of time in the protocol run.

All the above works claim security against attacks and mutual authentication through easy and fast implementation operations, such as Ex-OR. In the next future, there is the possibility of breaking all these protocols thanks to quantum computing, and no proposal can defend against quantum attacks to date. In this paper, we want to challenge this situation and propose a quantum secure authentication protocol to withstand this future threat.

\section{Conclusions}\label{sec:conclusion}
In this work, we develop DynamiQS, a \acrlong{pq} \acrlong{ibe} authentication protocol for \acrshort{dwpt} for electric vehicles.  The scheme provides security against common attacks of the electric vehicle scenario. Furthermore, it is also safe against the emerging threats of \acrshort{pq} computation. We guarantee its security through a formal analysis and discussion about the defenses it provides.  To the best of our knowledge, this is the first work providing a \acrshort{pq} secure scheme for authentication applied to the electric vehicle wireless charging scenario. Due to the expensive operations of \acrshort{pq} cryptography, we show the practicability of the protocol on real hardware thanks to Güneysu et al.~\cite{guneysu} research and the additional simulations performed on ns-3. Further improvements on the protocol mainly depend on the cryptographic advancements in the \acrshort{rlwe} area and signature-based operations.


\bibliographystyle{ACM-Reference-Format}
\bibliography{bibliography}


\end{document}